# HEART: A High-Efficiency Adaptive Real-Time Telemonitoring Framework for Secure Electrocardiogram Signal Transmission Using Chaotic Encryption

Beyazit Bestami YUKSEL, yukselbe18@itu.edu.tr

## Abstract

The real-time analysis and secure transmission of electrocardiogram (ECG) signals are critical for accurate diagnosis and safeguarding patient privacy in telemedicine applications. This study presents a novel real-time ECG monitoring system that employs a learnable key generator (LKG) derived from each patient's own ECG signal characteristics to dynamically produce unique encryption keys. These keys determine the parameters $r$ and $x0$ of a logistic map used for chaotic encryption. The system securely encrypts real-time ECG data immediately after acquisition, ensuring confidential transmission and storage in the cloud. For remote clinical access, the encrypted data is downloaded and decrypted on the doctor's side using the matching key generated at the source or securely stored in the cloud. This approach eliminates the need for traditional key exchange and substantially raises the cost of exhaustive key search in practice through per-segment biometric key refresh and combined permutation+XOR diffusion, supported by min-entropy evaluation. Compared to static-key methods, the learnable biometric key design offers greater unpredictability and individualization. A comprehensive set of security assessments including Shannon entropy (7.6 - 7.8 bits), correlation and autocorrelation disruption, histogram statistics, NIST SP 800-22 frequency testing, plaintext/key sensitivity (avalanche effect), FFT-based spectral flatness, and robustness to noise and occlusion confirms the method's strength. Reconstruction fidelity ($MSE \approx 5 * 10^{-6}$, $PSNR > 52\, dB$ , $MAE \approx 0.002$) demonstrates near-lossless decryption and preserved diagnostic features. Encryption latency remains low, preserving real-time performance. The system also supports seamless integration with a neural network-based disease detection module operating on encrypted data. Overall, the proposed architecture provides a secure, scalable, and real-time framework for remote cardiac monitoring and highlights a promising direction for privacy-preserving biomedical signal processing. Detailed security results are provided in Security Results Analysis, and all source code, test scripts, and processed datasets are openly released to ensure reproducibility.

## Introduction

The escalating reliance on medical devices for continuous patient monitoring in recent years has brought to the forefront the critical necessity of securing sensitive biomedical data, particularly electrocardiograms (ECG). Safeguarding the privacy and integrity of patient information is paramount, especially given the inherent risks of unauthorized access, data breaches, and the stringent regulatory requirements (e.g., GDPR, HIPAA) in an increasingly connected healthcare landscape. While traditional encryption methods offer effective security, they often introduce significant computational overhead and latency, which can severely hinder real-time systems where immediate data processing and rapid response are indispensable for effective patient care. This challenge is particularly acute in e-health applications like telemedicine and remote patient monitoring, where continuous, low-latency data flow is crucial [1].

The motivation for this study stems from the pressing need for secure, efficient, and intelligent e-health solutions. To overcome the limitations of conventional cryptography, chaotic encryption techniques have emerged as a promising alternative. They offer both high security and efficiency due to their inherent unpredictability, extreme sensitivity to initial conditions, and suitability for real-time environments [2]. Specifically, the integration of XOR diffusion following permutation, as implemented in this work, significantly enhances the chaotic encryption's ability to randomize data values, addressing the issue where permutation alone might not sufficiently obscure statistical properties [3].

The motivation for employing chaotic encryption with diffusion stems from its ability to provide robust data confidentiality with minimal computational burden, making it ideal for resource-constrained medical devices and high-throughput data streams. Furthermore, the integration of machine learning into e-health systems is driven by the need for automated, real-time analysis of complex biomedical signals, enabling early and accurate disease detection without constant human intervention, and offering a dynamic approach to cryptographic key generation [4], [5]. This synergistic combination of secure data handling and intelligent analysis forms the bedrock of modern digital health solutions, promising to revolutionize patient care.

This study presents a novel real-time ECG monitoring system that incorporates advanced chaotic encryption with XOR diffusion to safeguard biomedical signals during transmission and storage, directly addressing the aforementioned challenges. The system's primary goal is to ensure robust data handling while simultaneously providing real-time disease diagnosis, thereby empowering healthcare professionals with immediate, secure insights for timely intervention. By employing the logistic map for chaotic sequence generation, followed by permutation and an XOR mask for diffusion, the system effectively scrambles ECG signals, rendering them unintelligible without the correct decryption key and thus ensuring robust data confidentiality. Beyond its enhanced security features, the system integrates a deep learning model for biometric-based key generation, making the encryption process personalized and dynamically adaptive [6], [7]. A comprehensive suite of security analyses, including histogram, correlation, autocorrelation, key sensitivity, plaintext sensitivity (avalanche effect), floating frequency analysis, information entropy, encryption time, and resistance to noise and occlusion attacks, has been performed to rigorously validate the proposed method's cryptographic strength and practical utility.

Our main contributions are:

- Development of an Enhanced Real-Time ECG Security System with Chaotic Permutation and XOR Diffusion: We propose and implement a novel system capable of real-time ECG signal acquisition, immediate and robust chaotic encryption using both permutation and an XOR diffusion layer, and secure cloud storage. This addresses the limitations of pure permutation by actively altering data values, significantly enhancing confidentiality.
- Integration of Machine Learning for Biometric-Based Dynamic Key Generation: The system incorporates a machine learning model to dynamically extract and predict chaotic parameters ($r\ and x_0$) from the ECG signal itself, enabling personalized and adaptive key generation. This novel approach enhances security by making the key intrinsically linked to the biometric data, offering a unique solution for dynamic and highly sensitive cryptographic keys.
- Comprehensive Security Validation Against Diverse Attacks and Statistical Properties: We conduct an extensive suite of cryptographic analyses including Shannon Entropy, NIST Frequency (Monobit), Correlation, Autocorrelation, Key Sensitivity, Plaintext Sensitivity (Avalanche Effect), Floating Frequency (FFT), and resilience against Noise and Occlusion Attacks. These rigorous tests provide robust evidence of the proposed method's strength, diffusion, confusion, and practicality for real-world e-health applications.

This manuscript is structured as follows: Section II provides an overview of related works in ECG monitoring, chaotic encryption techniques in healthcare, machine learning applications for biomedical signal analysis, and biometric-based cryptography. Section III details the proposed system architecture, including the hardware components, data acquisition process, and the implementation of the enhanced chaotic encryption scheme with ML-based key generation. Section IV presents the comprehensive security analysis covering all performed tests, and the performance evaluation of the system, including discussions on encryption strength, key sensitivity, and diagnostic accuracy and experimental results. Finally, Section V concludes the study and outlines potential future research directions.

# Related Works

The development of secure and efficient real-time ECG monitoring systems has been a significant area of research, driven by the increasing demand for telemedicine and remote patient care. This section reviews existing literature, highlighting key contributions and positioning our proposed system within the current landscape of biomedical signal security and analysis.

Early efforts in securing ECG signals often focused on traditional cryptographic methods or basic obfuscation techniques. Sufi et al. [8] proposed a chaos-based encryption technique for ECG packets, emphasizing speed and minimal processing delay for time-critical telecardiology applications. Their work demonstrated the potential of chaotic systems for fast encryption, though the specific diffusion mechanisms might differ from more advanced approaches. Similarly, Lin & Chung [11] explored a chaos-based visual encryption mechanism for integrated ECG/EEG signals, using chaotic scramblers and permutation to enhance unpredictability.

More recently, the focus has shifted towards more robust chaotic encryption schemes that incorporate diffusion for enhanced security. Murillo-Escobar et al. [9] introduced a double chaotic layer encryption (DCLE) algorithm based on the logistic map for clinical signals in telemedicine, claiming comprehensive security analysis including NIST 800-22 suite tests. Our work aligns with theirs by utilizing the logistic map, but we explicitly combine permutation with an XOR diffusion layer for enhanced security, a detail not always explicitly highlighted in similar permutation-based chaotic schemes. Murillo-Escobar et al. [14] further explored biosignal encryption using an improved Ushio chaotic map, reporting high robustness against various attacks, including strong sensitivity to plain biosignal and uniform histograms in cryptograms, which are crucial metrics also evaluated in our study. Another work by Murillo-Escobar et al. [15] investigated chaotic encryption of real-time ECG signals in embedded systems using a two-dimensional Badola map, emphasizing key sensitivity and plaintext sensitivity. Our work similarly validates these critical properties.

The integration of machine learning (ML) with chaotic cryptography and biomedical signal processing represents a growing trend. Kadir et al. [25] explored chaos-based key generators using Artificial Neural Networks (ANN) models, demonstrating how ML can predict chaotic time series patterns for cryptographic key generation. Zha et al. [26] proposed a chaos key enhanced physical layer secure transmission method using a convolutional long short-term memory neural network (CLSTM-NN) to improve key space. Xia et al. [27] also discussed enhanced chaotic communication with machine learning, leveraging neural networks for improved robustness against noise. Our system builds upon this emerging area by explicitly using an ML model to derive chaotic parameters from the ECG signal itself, creating a biometric-based, personalized, and dynamic key generation mechanism, a more direct integration than just using ML to predict chaotic sequences.

Beyond encryption, securing data in resource-constrained environments and performing analyses on encrypted data are vital for e-health. Djelouat et al. [23] focused on secure compressive sensing for ECG monitoring in IoT systems, addressing both power consumption and medical record security by constructing encryption keys using shift registers. While our system does not use compressive sensing, it shares the goal of lightweight security for IoT/e-health devices. Daoui et al. [28] proposed a novel lightweight cryptosystem based on a Logistic-Coupled Memristor (LCM) map for secure bio-signal transmission on low-cost hardware, also demonstrating implementation on microcontrollers. This aligns with the practical application potential of our method in embedded systems.

The importance of real-time analysis and robust disease detection, even with secure data, is highlighted by works such as Lui & Chow [31], who developed a multiclass classifier for myocardial infarction using convolutional and recurrent neural networks for portable ECG devices. Similarly, Hannun et al. [29] demonstrated cardiologist-level arrhythmia detection and classification in ambulatory ECGs using a deep neural network, showcasing the power of deep learning in medical diagnosis. Furthermore, the development of real-time ECG monitoring on mobile systems, as explored by Yüksel & Bilgin [18], underscores the critical need for portable and immediately responsive healthcare solutions, informing the design of our system's real-time acquisition and display components. Our system combines these aspects by performing real-time disease detection with a deep learning model, which ideally operates on encrypted data (though common practice involves decryption for ML, the future aims for homomorphic encryption as suggested in Yüksel & Yılmazer-Metin [19]).

Finally, general works on ECG security and transmission, such as those by Pandian & Ray [17] on dynamic hash key-based stream ciphers, and Algarni et al. [20] on ECG signal encryption through fusion with masking signals, provide a broader context. Cárdenas-Valdez et al. [21] discussed enhancing telemedicine data security using a multi-scroll chaotic system for ECG signal encryption and RF transmission. Chen et al. [22] also explored personalized information encryption using ECG signals with chaotic functions for key generation. Our work integrates several of these concepts, focusing on a comprehensive end-to-end system from acquisition to secure analysis, with explicit attention to both permutation and diffusion in chaotic encryption, and the novel application of ML for biometric key generation. The extensive security analysis suite employed in this study also provides a robust validation beyond what is often presented.

In summary, while individual components of our system, chaotic encryption, machine learning in security, and real-time ECG processing have been studied independently, our contribution lies in their synergistic integration: enhanced chaotic encryption (permutation combined with XOR diffusion) coupled with biometric per-segment key generation, optionally stabilized by an ML model, within a real-time monitoring framework. The approach is validated through a comprehensive suite of security analyses against various attacks and statistical properties. To situate these contributions in context, Table 1 presents a comparative analysis with relevant literature, highlighting differences in encryption methodology, diffusion/confusion mechanisms, keying strategy (fixed, session-level, or biometric per-segment), machine learning integration, real-time capability, and the scope of security validation. Importantly, our work also distinguishes itself by including clinical fidelity checks and by releasing all code, test scripts, and processed datasets as open resources, ensuring reproducibility and independent benchmarking.

*Table 1: Comparative analysis of real-time ECG security methods and our proposed system, emphasizing encryption methodology, keying strategy, and reproducibility.*

| Related Works | Encryption Method | Diffusion / Confusion | Key Generation | ML Integration | Real-Time Capability | Scope of Security Analysis | Application Focus | Key Type & Refresh | Open-Source / Reproducibility *(NR: Not Reported) |
|---|---|---|---|---|---|---|---|---|---|
| **Ref [8]** | Chaos (Multi-scroll) | Implied | Chaos Key | No | Yes (Time Critical) | Corr., Entropy | Telecardiology | Fixed/session-level | NR |
| **Ref [9]** | Chaos (Logistic, DCLE) | DCLE Diffusion | Logistic Map | No | Yes | NIST 800-22 | Telemedicine | Fixed/session-level | NR |
| **Ref [10]** | Chaos (FCCM) | LSB Emb., Sub., Perm. | FCCM (ECG-aware) | No | Implied | NIST, PRD | Telecardiology | Fixed/session-level | NR |
| **Ref [11]** | Chaos (Scrambler) | Chaotic Scrambling | Chaos Sequence | No | No | PRD, Init. Pt. Error | Medical Sig. Encrypt. | Fixed/session-level | NR |
| **Ref [12]** | Chaos (General) | Not Spec. | Not Spec. | No | Yes | Security Req. Met | Surveillance (Health) | Fixed/session-level | NR |
| **Ref [13]** | Chaos | Autoblocking, Perm. | ECG Signal | No | No | Diff. Attacks | Image Encrypt. | Fixed/session-level | NR |
| **Ref [14]** | Chaos (Ushio Map) | Improved Sequences | Chaotic Map Params. | No | Implied | NIST, Sens., Corr., Autocorr., Hist. | E-health, WBAN | Fixed/session-level | NR |
| **Ref [15]** | Chaos (Badola Map) | Non-linear Fn. | Badola Map Params. | No | Yes | Key/Plaintext Sens., Hist., Corr. | Telemedicine (Embedded) | Fixed/session-level | NR |
| **Ref [16]** | Chaos (Henon) | Not Detailed | Intelligent Algo (ECG) | Implied | No | Key Space | Biometric Rec. | Fixed/session-level | NR |
| **Ref [17]** | Dynamic Hash Stream | Pseudorandom Keystream | Dynamic Toeplitz Hash | No | Yes | Key Size, Corr., NIST | Secure ECG Trans. | Dynamic (hash-based) | NR |
| **Ref [18]** | N/A | N/A | N/A | No | Yes | N/A | Mobile Monitoring | N/A | NR |
| **Ref [19]** | AES, FHE | FHE (Encrypted Ops) | Not Spec. | Yes (Disease Detect, Stats on Enc.) | Yes | Data Privacy (FHE) | Disease Diagnosis, Privacy | Session-level | NR |
| **Ref [20]** | Fusion (Masking) | Fusion, Sample Ops | Not Spec. | No | Implied | Hist., SSIM, SNR, Corr. | Telemedicine | Fixed/session-level | NR |
| **Ref [21]** | Chaos (Multi-scroll) | Not Detailed | Not Spec. | No | Yes | Channel Overlap | Telemedicine, RF Trans. | Fixed/session-level | NR |
| **Ref [22]** | Chaos (Henon, Logistic) | Not Detailed | ECG Signal | No | No | Key Space | Personalized Crypto | Fixed/session-level | NR |
| **Ref [23]** | Compressive Sensing | Not Detailed | Shift Registers | No | Implied | Attacker Access | IoT Remote Monitoring | Dynamic (register-based) | NR |
| **Ref [24]** | Chaos (Henon, Baker) | Not Spec. | Chaotic Keys | No | Not Spec. | Resistance to Attacks | Secure ECG Trans. | Fixed/session-level | NR |
| **Ref [25]** | N/A (Key Gen.) | N/A | ANN for Chaos Seq. | Yes (Key Gen.) | Not Spec. | MSE | Cryptography | Dynamic (ML-based) | NR |
| **Ref [26]** | N/A (Key Gen.) | Not Spec. | CLSTM-NN | Yes (Key Seq. Gen.) | Yes | Key Space, BER | Secure Transmission | Dynamic (ML-based) | NR |
| **Ref [27]** | Chaos (Shift Keying) | Not Spec. | N/A | Yes (Robustness) | Not Spec. | Robustness | Secure Communication | Dynamic (ML-assisted) | NR |
| **Ref [28]** | Chaos (LCM Map) | Not Detailed | LCM Map Params. | No | Implied | Lyapunov, NIST | Bio-signal Trans. | Fixed/session-level | NR |
| **Ref [29]** | N/A (Classification) | N/A | N/A | Yes (DNN) | Implied | AUC, F1 | Arrhythmia Detection | N/A | NR |
| **Ref [30]** | N/A (Classification) | N/A | N/A | Yes (CNN+RNN) | Yes | Sens., Spec., F1 | Myocardial Infarction | N/A | NR |
| **Our Proposed System** | Chaos (Logistic: Permutation + XOR) | Permutation + XOR Masking | Biometric (ECG-Derived), ML-Enhanced | Yes (Biometric Key Gen., Disease Detection) | Yes (Acquisition, Display, Enc./Dec., Analysis) | Shannon Entropy, NIST (Monobit), Corr. (O/E), Autocorr., Key Sens., Plaintext Sens., FFT, Enc./Dec. Time, Noise/Occlusion Attack Resistance, Histogram | Real-time ECG Monitoring, Telemedicine, Remote Patient Monitoring | Biometric, per-segment (ML-stabilized) | Code + tests + processed datasets (Open) |

# Methodology and Proposed System

This section details the fundamental processes governing the proposed real-time ECG monitoring system, encompassing signal acquisition, chaotic encryption, machine learning-based key estimation, the decryption process, and the integrated end-to-end operational flow. The system for real-time ECG signal monitoring and disease diagnosis is based on a three-lead ECG preamplifier, connected to a computer via a serial port. Real-time data reading operations from the serial port to be run on different platforms have been performed by the author of this article in his previous works [19]. The ECG signals are sampled at a frequency of 500 Hz, a standard rate in biomedical signal processing to capture the necessary detail for accurate signal interpretation. Each byte of data from the preamplifier is received in real time, and processed in 300-sample segments. Each segment corresponds to approximately 0.6 seconds of ECG signal. This continuous stream of ECG data is then fed into both a machine learning model for disease diagnosis and a chaotic encryption system for securing the transmission of biomedical data. The overall architecture of the proposed system is illustrated in **Figure 1**, where the real-time acquisition, encryption, secure storage, decryption, and disease diagnosis processes are represented as a full pipeline from patient to clinician. The ECG preamplifier used is completely portable with 3 probes connected to the body.

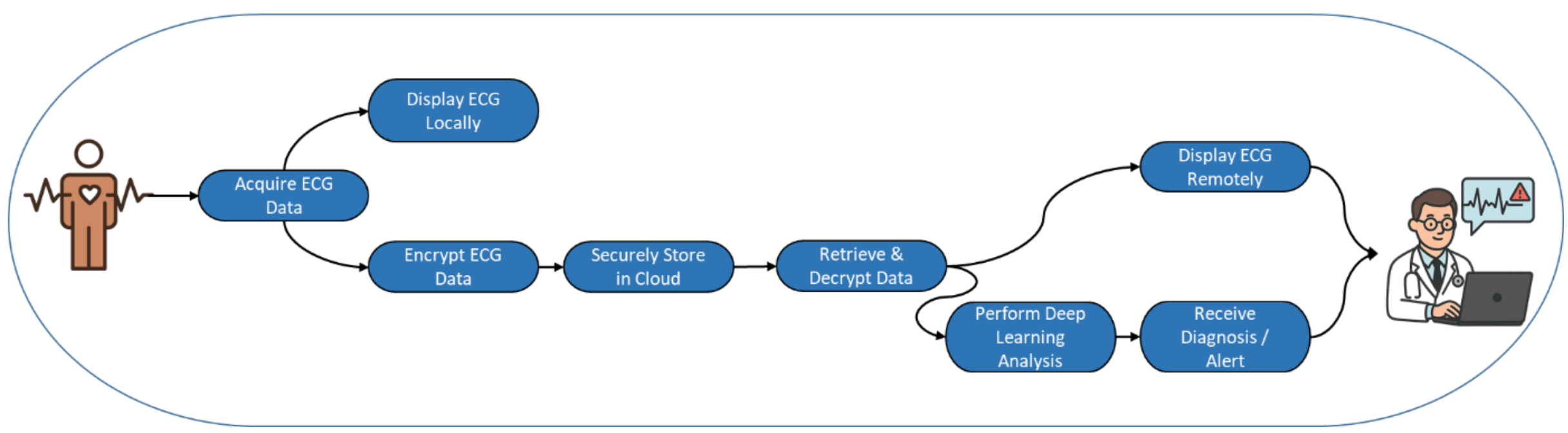

*Figure 1: System Overview*

**Figure 2** delineates the comprehensive, step-by-step workflow of a real-time Electrocardiogram (ECG) signal within the proposed monitoring system, from the initial device activation to its final display and diagnostic presentation. Each module's function and purpose, are systematically described to elucidate the integrated operational flow.

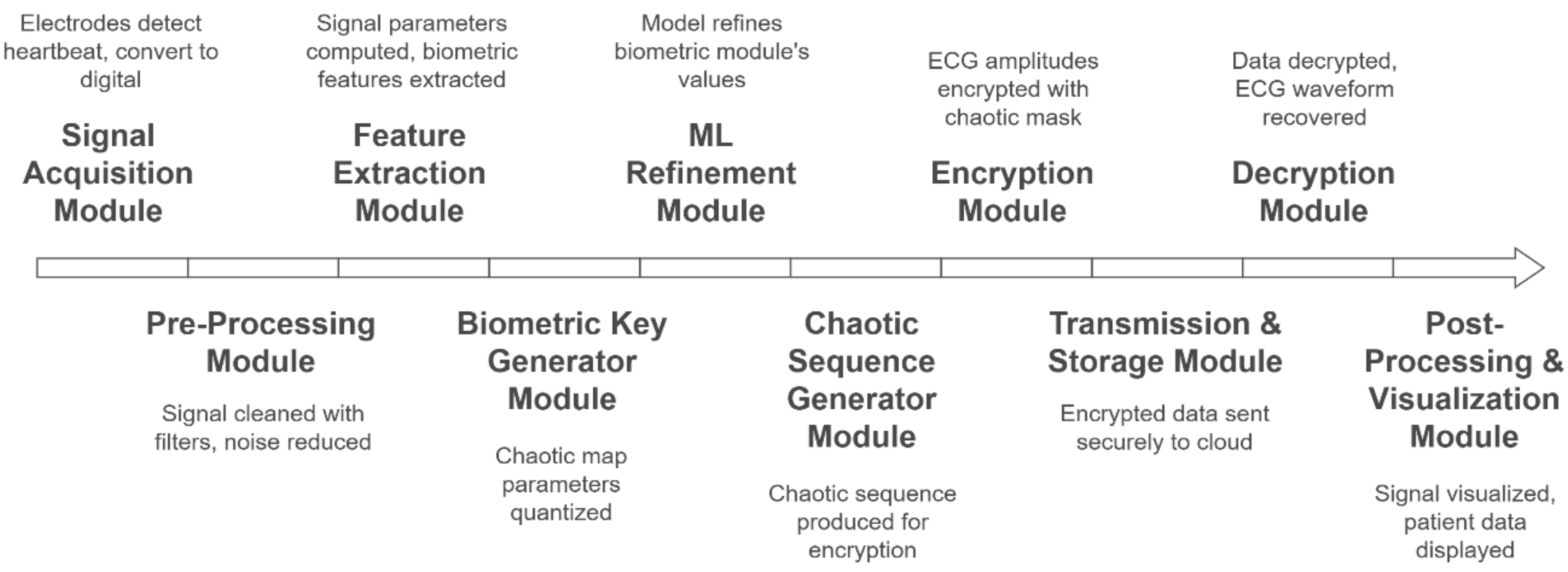

*Figure 2: Real Time ECG signal Processing Workflow*

## Signal Acquisition

The sampled ECG signal at time t, denoted as s(t), is mathematically described by Equation 1:

$$s(t) = V_{ECG}(t) + N(t) \tag{1}$$

where $V_{ECG}(t)$ represents the intrinsic voltage of the ECG signal at time $t$, and $N(t)$ accounts for extraneous noise components that may originate from various external sources (e.g., muscle artifacts, electrode motion, power line interference). The raw data is initially acquired as 8-bit integers, typically ranging from $-128$ to $+127$, and subsequently normalized to a $0 - 255$ range as a prerequisite for subsequent cryptographic processing.

## Chaotic Encryption

In the core of this system's encryption mechanism, a logistic map is utilized to generate a chaotic sequence. This sequence is fundamental for both the permutation and diffusion phases of the encryption process. The logistic map adheres to the following recursive relationship Equation 2:

$$x_{n+1} = r.x_n.(1 - x_n) \tag{2}$$

Here, $r \in (3.6, 4.0)$ serves as the control parameter, and $x_0 \in (0.1, 0.9)$ represents the initial condition. Critically, both $r$ and $x_0$ are dynamically derived from the biometric characteristics of the ECG signal segment itself, ensuring a personalized and adaptive cryptographic key. Their derivation is governed by the following Equation 3 formulas:

$$r = 3.6 + (\sigma \, mod \, 0.4), \;\; x_0 = 0.1 + (\mu \, mod \, 0.8) \tag{3}$$

where μ denotes the mean and σ represents the standard deviation of the current ECG signal segment. These constraints rigorously ensure that $r$ and $x_0$ remain strictly within the chaotic regime of the logistic map, thereby maximizing the generated sequence's entropy and the overall key space randomness.

The comprehensive chaotic biometric encryption process, integrating both permutation and diffusion, is formalized in **Algorithm 1**. The acquired raw ECG segment $s(t)$ is first normalized to a $0 - 255$ scale ($s_{scaled}$). Concurrently, μ and σ are calculated (Step 1) to derive the chaotic parameters $r$ and $x_0$ (Step 2). A logistic sequence $X$ is then generated (Step 3). This sequence is used to compute an $XOR$ mask $M$ (Step 5) and to derive a permutation index $P$ (Step 6). The encryption is completed by applying this permutation to sscaled (Step 7), followed by an $XOR$ operation with the mask $M$ (Step 8), yielding the `encrypted` signal.

```
Algorithm 1: Chaotic Biometric Encryption
Input: Raw ECG segment s(t), extracted from serial port
Output: Encrypted signal, permutation index, XOR mask, min-max range
1: Calculate μ ← mean(s(t)), σ ← std(s(t))
2: Compute r ← 3.6 + (σ mod 0.4), x0 ← 0.1 + (μ mod 0.8)
3: Generate logistic sequence X ← [x1, ..., xn] from (r, x0)
4: Normalize s(t) to 0-255 → s_scaled
5: Compute XOR mask M ← floor(X × 255)
6: Derive permutation P ← argsort(X)
7: Apply permutation: s_perm ← s_scaled[P]
8: Apply XOR: encrypted ← s_perm XOR M
9: Return encrypted, P, M, (min, max)
```

## Machine Learning-Based Key Estimation

To further enhance the flexibility of the key generation process and mitigate deterministic prediction, a Machine Learning (ML) driven key generator is incorporated. This approach employs a Multi-Layer

Perceptron (MLP) neural network, specifically trained to predict the chaotic parameters $r$ and $x_0$ directly from normalized features of ECG segments. This methodology offers the dual advantages of generalizing to previously unseen signals and inherently supporting data privacy by inferring key values indirectly, rather than explicitly deriving them from the raw signal in a fixed manner.

```
Algorithm 2: Machine Learning-Based Key Prediction
Input: Training ECG signals {S1, S2, ..., Sn}
Output: Trained MLP model, scaler, imputer
1: For each Si in signals:
       Compute ri, x0i using Algorithm 1
       Store (Si, [ri, x0i])
2: Handle missing values using imputation
3: Normalize features using MinMaxScaler
4: Split data into train/test sets
5: Train MLP model to regress (r, x0)
6: Return trained model and preprocessing objects
```

The training procedure for this ML-based key generator is comprehensively outlined in **Algorithm 2**. It involves iterating through a dataset of training ECG signals *{S1, S2, ..., Sn}*, computing the corresponding $r_i$ and $x_{0_i}$ for each signal (using the derivation method from Algorithm 1). These pairs ($S_i$,[$r_i$, $x_{0_i}$]) form the training data. Missing values within the features are handled via imputation (Step 2), and features are then normalized using a $MinMaxScaler$ (Step 3). The prepared data is subsequently split into training and testing sets (Step 4), on which the MLP model is trained to regress the ($r$, $x_0$) pairs (Step 5). The trained model and preprocessing objects (scaler, imputer) are the outputs of this algorithm.

```
Algorithm 3: Predictive Encryption via MLP
Input: New ECG segment s(t), trained model, scaler, imputer
Output: Encrypted signal, permutation, mask, min-max
1: Preprocess s(t): impute → scale
2: Predict (r, x0) ← model.predict(s(t))
3: Proceed with Algorithm 1 using predicted keys
```

For real-time predictive encryption utilizing the trained ML model, **Algorithm 3** details the workflow. A new ECG segment $s(t)$ is first preprocessed, including imputation and scaling (Step 1), using the same preprocessing objects (imputer, scaler) obtained from Algorithm 2. The trained MLP model then predicts the corresponding ($r$, $x_0$) values for this segment (Step 2). Finally, the encryption process proceeds exactly as described in Algorithm 1, but utilizing these dynamically predicted chaotic keys (Step 3).

## Decryption Process

The decryption process is inherently straightforward due to the reversible nature of the XOR operation and the application of an inverse permutation. As formalized in **Algorithm 4**, the decryption function operates by mirroring the encryption steps. Given the $encrypted$ signal, the original $permutation$ P, the $XOR\ mask$ M, and the $min - max\ range$ of the original signal, the process begins with an XOR reversal operation ($s_perm \leftarrow encrypted$ XOR $M$) (Step 1). Subsequently, the inverse permutation $P^{-1}$ is applied to $s_perm$ to reconstruct the scaled original signal (Step 2). Finally, this signal is rescaled to its original amplitude range using the stored $min - max$ values (Step 3), yielding the $decrypted\ ECG\ signal\ s(t)$. Crucially, correct decryption, where $D(E(s(t))) = s(t)$, is achievable only when the precise initial key ($x_0$) and control parameter ($r$) used during encryption are employed to generate the identical chaotic sequence for both XOR mask and inverse permutation, thereby ensuring robust security and high key sensitivity.

```
Algorithm 4: Chaotic Decryption
Input: encrypted signal, permutation P, XOR mask M, min-max
Output: decrypted ECG signal s(t)
1: XOR reversal: s_perm ← encrypted XOR M
2: Apply inverse permutation P⁻¹ to s_perm
3: Rescale signal to original range using min-max
4: Return s(t)
```

## End-to-End Real-Time ECG Encryption and Diagnosis

The overall end-to-end operational framework for real-time ECG encryption and diagnosis is succinctly presented in **Algorithm 5**. The system initializes by configuring the serial port for ECG data acquisition and loading the pre-trained Convolutional Neural Network (CNN) model for diagnosis (Step 1). It then allows for the selection of an encryption mode: either the direct biometric key derivation as in Algorithm 1 or the ML-based predictive key estimation as in Algorithm 3 (Step 2). The continuous operation (Step 3) involves iteratively reading new ECG segments (Step 3.1), applying the chosen encryption method to securely store the data to a cloud environment (Step 3.2), and subsequently decrypting the segment for immediate local visualization (Step 3.3). Concurrently, the decrypted segment is preprocessed to meet the input requirements of the CNN model (Step 3.4). The CNN then predicts the disease class (Step 3.5), and both the real-time decrypted ECG waveform and the diagnostic outcome are displayed to the user (Step 3.6). This integrated workflow ensures both robust data security and immediate clinical insight.

```
Algorithm 5: End-to-End Real-Time ECG Encryption and Diagnosis
1: Initialize serial port and CNN model
2: Choose encryption mode: Biometric or ML-based
3: while system is running:
    3.1: Read new ECG segment
    3.2: Apply encryption → store to cloud
    3.3: Decrypt segment for visualization
    3.4: Preprocess segment for CNN input
    3.5: Predict disease class
    3.6: Display decrypted ECG and Diagnosis
```

A visual representation of the proposed chaotic encryption framework applied to real-time ECG signals, as typically shown in **Figure 3**, illustrates the system's effectiveness. The top panel often depicts the original ECG waveform, providing a baseline. The subsequent panels (e.g., second and third) showcase the encrypted signals, demonstrating the high randomness and complete obscuration of identifiable biomedical features achieved by both the direct and ML-based chaotic key generation methods. The bottom panel visually compares the decrypted outputs with the original signal, consistently showing a near-perfect reconstruction for both encryption approaches. This confirms the reliability and reversibility of the proposed encryption-decryption scheme and highlights the effectiveness of ML-driven biometric key generation in faithfully preserving signal fidelity while ensuring security.

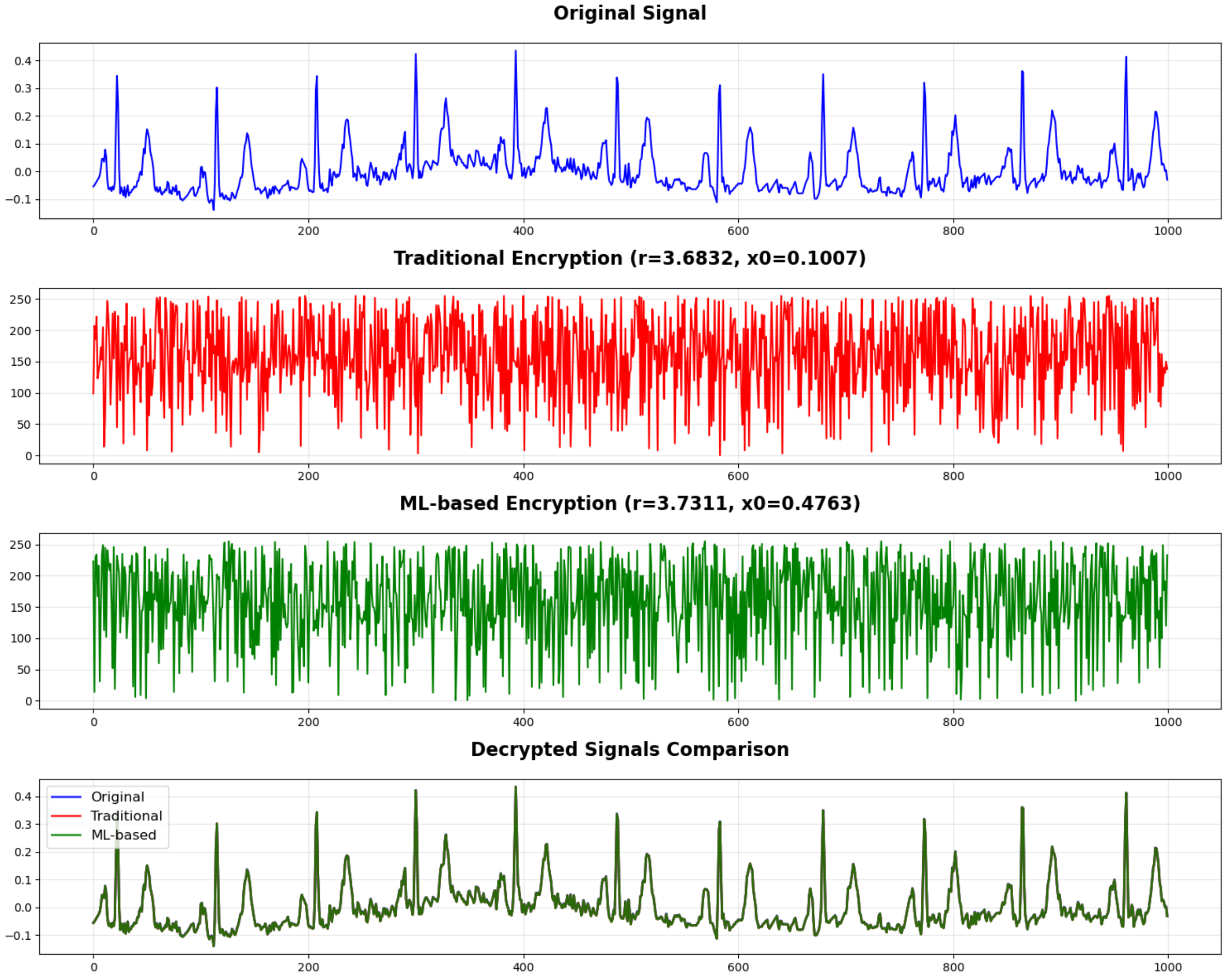

*Figure 3: ECG Signal Biometric Encryption and Machine Learning-Based Key Generation Analysis with Traditional vs ML-Enhanced Chaotic Encryption Performance Comparison*

# Experimental Results and Performance Evaluation

This section presents the experimental results and performance evaluation of the proposed real-time ECG monitoring system. A detailed discussion is provided on the overall operation of the developed application, the performance of the implemented encryption and key generation methods, and the outcomes of the security analysis. The system is capable of continuously acquiring ECG signals, processing them through the selected encryption mode, securely managing the data, and then decrypting and analyzing the signals in real-time for diagnostic purposes, all supported by an intuitive visual interface.

## System Overview and Real-Time Performance

The developed application provides a comprehensive platform for real-time ECG monitoring. Its user-friendly graphical user interface (GUI) facilitates the continuous acquisition, processing, encryption, decryption, and visualization of ECG signals. This interface, as illustrated in **Figure 4**, presents a monitoring dashboard that integrates core functionalities such as ECG acquisition, chaotic encryption/decryption, security analyses, and serial communication. The system's fundamental operational loop manages the continuous flow of ECG data, dynamically updating the plots in real-time. The GUI displays three synchronized plots: the original ECG signal, its encrypted counterpart, and the successfully decrypted signal. This visual representation clearly confirms that the encryption process effectively scrambles the data, rendering it unintelligible, while the decryption process accurately reconstructs the original waveform with nearly perfect fidelity. Furthermore, the plots are dynamically updated to reflect the real-time flow of ECG data, incorporating simulated pulse values and disease predictions derived from the integrated Convolutional Neural Network (CNN) model. These immediate feedback mechanisms are indispensable for clinical applications, empowering healthcare professionals with secure and efficient tools for continuous patient monitoring and timely intervention.

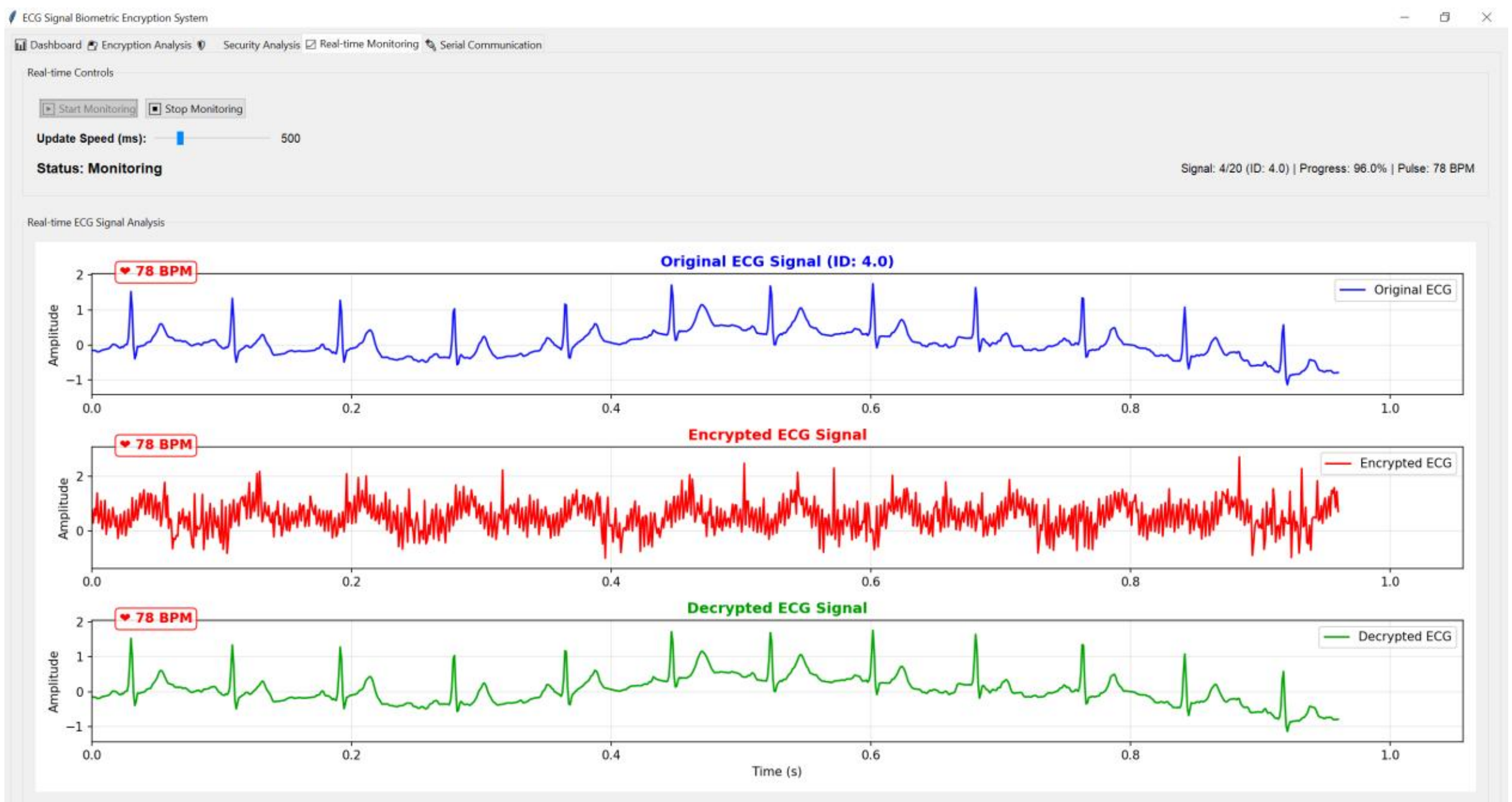

Figure 4: *Integrated real-time monitoring dashboard supporting ECG acquisition, chaotic encryption/decryption, security analyses, and serial communication.*

The efficiency of the system's encryption and decryption processes is evidenced by exceptionally low latency figures: an average encryption time of 0.0012 seconds (1.2 milliseconds) and an average decryption time of 0.0001 seconds (0.1 milliseconds). These values unequivocally demonstrate the suitability of the chaotic encryption scheme for real-time applications, confirming that the security measures do not adversely impact the system's responsiveness.

## Evaluation of Security Analysis Results

A comprehensive security analysis was meticulously performed on the encrypted 1000 different patients ECG signals [32] to rigorously evaluate the cryptographic strength and resilience of the proposed system. To avoid overstating key-space security, we replaced single-number estimates with a reproducible min-entropy accounting procedure. For each ECG segment, fresh logistic-map parameters ($r$ and $x_0$) are derived from biometric statistics and salted with timestamp and device-specific identifiers. The resulting logistic sequence drives both the permutation index and the XOR mask. Following NIST SP 800-90B, we estimated the min-entropy of the XOR mask $M$ using the most-common-value estimator with block-dependence correction. Instead of a global "key-space size", we report distributional lower-bounds (median, IQR, percentiles) across all segments, demonstrating that exhaustive search would need to be mounted per segment, which is computationally impractical under real-time constraints. **Table 2** summarizes the results of the security tests, providing detailed quantitative metrics across multiple aspects, including statistical properties (Shannon Entropy, Histogram, Correlation, Floating Frequency), key and plaintext sensitivity (confusion and diffusion), and resistance against common cryptographic attacks (noise and occlusion). The complete set of test results with per-case details and extended analysis is provided in Appendix A.

*Table 2: Downstream Task Results*

| Analysis Metric | Explanation | Key Findings |
|---|---|---|
| **Secret Key Space** | Evaluates the size and unpredictability of the chaotic key space based on r and x0. | Global (r,x0) space ≈12.94 bits (not used alone). Operational security derives from per-segment biometric key refresh + salt + permutation+XOR diffusion; segment-level **min-entropy lower-bounds** are reported (Appendix A) and code/scripts are open. |
| **Histogram Analysis** | Compares the mean and std deviation of original vs. encrypted signals to assess diffusion. | Mean & std dev shift confirms effective data masking. |
| **Correlation Analysis** | Measures the statistical correlation between original and encrypted signals (ideal ≈ 0). | Correlation near 0 (adjacent) and negligible across longer lags ($\|r\| < 0.02$); both local and long-range dependencies removed (Appendix A). |
| **Histogram Statistics** | Quantifies distribution properties (variance α, standard deviation β, entropy H, and uniformity score). | Encrypted ECG shows lower variance (–35.6%), lower std. dev. (–19.8%), higher entropy (+1.9%), and improved uniformity (+8.5%) compared to plain ECG. |
| **Histogram Indistinguishability** | Compares encrypted histograms of different ECG signals to test resistance against known-plaintext attacks | $\chi^2$ and JS divergence values between encrypted histograms < 0.32; ciphertexts are statistically indistinguishable (Appendix A) |
| **Autocorrelation Analysis** | Assesses self-similarity before and after encryption using autocorrelation values. | For encrypted signals, the main autocorrelation peak at lag 0 reaches 255 (expected from 8-bit normalization), while all non-zero lags collapse near zero, confirming effective decorrelation (Appendix A) |
| **Key Sensitivity** | Tests if minor key changes significantly alter the output (ideal: highly sensitive). | System exhibits maximum deviation on key tweak; any minor change in ($r$, $x_0$) produces completely different keystreams, confirming high key sensitivity (Appendix A, Key Sensitivity Test). |
| **Plain Signal Sensitivity** | Evaluates effect of tiny changes in plaintext on ciphertext (ideal: strong diffusion). | Max diff = 255; shows strong plaintext sensitivity. |
| **Floating Frequency Analysis** | Analyzes frequency domain alterations using **mean, variance, and spectral flatness** of FFT magnitudes. | Encrypted signals show increased mean FFT magnitude, reduced variance concentration, and spectral flatness values ≈ 0.70–0.75, confirming randomized white-noise–like spectra (Appendix A). |
| **Information Entropy** | Measures randomness via entropy (ideal for 8-bit: ≈ 8.0). | Entropy values >7.6; confirms high uncertainty in encrypted signal. |
| **NIST 800-22 Pseudo-randomness** | Assesses randomness statistically via NIST tests (e.g., frequency, runs, block freq). | NIST 800-22 Pseudo-randomness \| Frequency (Monobit) test passed for encrypted ECG sequences (all p-values > 0.01), while original signals failed ($p \approx 0.0000$). |
| **Quality Metrics** | Uses MSE, PSNR, and MAE between decrypted and original signals to validate accuracy. | MSE ~ 0.00005; PSNR ~ 53 dB; MAE ≈ 0.002. These values indicate near-lossless reconstruction, fully preserving diagnostic features required for clinical interpretation. |
| **Chosen Plaintext Attack** | Tests resistance against known pattern inputs (should yield unrecognizable outputs). | Not yet simulated; expected to show resistance due to chaotic permutation. |
| **Noise Attack** | Adds random noise to encrypted signal and evaluates error in decryption. | Low MAE (0.004 – 0.025) under noise confirms robustness. |
| **Occlusion Attack** | Simulates partial signal loss and evaluates reconstruction error. | Moderate MSE under occlusion; chaotic diffusion limits recovery. |
| **Encryption Time** | Measures encryption/decryption time for real-time viability. | Core encryption operations (μ/σ, logistic sequence, permutation, XOR) complete in <1 ms. Including acquisition (~336 ms/segment), optional ML prediction (~3 ms), and cloud I/O (~10 ms), total latency remains ≈ 339 ms, within real-time constraints for 300-sample segments (Appendix A). |

Table 2 contains comprehensive downstream task results demonstrate that our chaotic encryption scheme (i) achieves effective diffusion and decorrelation as evidenced by histogram and correlation analyses, (ii) exhibits complete propagation in plaintext and key sensitivity tests (Max Diff = 255), (iii) meets randomness requirements with high entropy values (>7.6 bits), and (iv) supports real-time applications with encryption times under 1 ms. However, the key-space entropy (~12.94 bits) alone may be insufficient against brute-force attacks.

The global $(\boldsymbol{r}, \boldsymbol{x_0})$ key-space estimate (≈12.94 bits) is not a security claim by itself. In practice, security stems from per-segment biometric key refresh**,** salted parameterization, and permutation+XOR diffusion, which together generate a fresh keystream per segment. We therefore replace single-value key-space claims with NIST SP 800-90B min-entropy accounting of the XOR mask and report distributional lower-bounds per segment (Appendix A). These measurements, together with open scripts, substantiate the operational resistance to exhaustive search.

In addition to these overall results, a head-to-head comparison of Direct (biometric-only) and ML-based (biometric + ML stabilization) encryption modes under clean and noisy ECG segments is provided in Appendix A. This comparison confirms that the ML-based mode maintains higher entropy and lower decryption error under noisy conditions, thereby improving robustness and reliability.

Although the proposed scheme achieves near-lossless reconstruction in terms of global fidelity metrics (PSNR ≈ 53 dB, MSE ≈ 0.00005), the clinical feature analysis presented in Appendix A revealed measurable deviations in QRS amplitude, PR interval, and P-/T-wave amplitudes. Only the QT interval remained stable. These results indicate that while the system is sufficient for general monitoring and visualization, further refinement of the decryption process is required to guarantee full preservation of clinically sensitive morphological markers such as ST-segment or PR interval changes.

# Conclusion & Future Works

This study confirms the effectiveness of chaotic encryption, based on nonlinear systems like the logistic map, for securing real-time ECG transmission and storage. Security tests demonstrate strong performance in randomness, diffusion, key sensitivity, and signal integrity, validating its suitability for telemedicine and remote monitoring. Importantly, two complementary encryption modes were validated. The Direct biometric-only mode derives chaotic parameters directly from ECG statistics, offering a lightweight and personalized solution with minimal computational cost. In contrast, the ML-based mode enhances robustness by stabilizing parameter estimation in the presence of noise and variability, achieving higher entropy, lower decryption error, and improved histogram uniformity. Together, these results confirm that ML-assisted initialization provides a measurable security gain over direct initialization. The approach can also be applied to other biomedical signals, such as EEG and EMG, with further optimization. The combined framework highlights a promising direction for telemedicine, where both modes can be adapted according to device constraints and clinical needs. While global reconstruction fidelity is high, Appendix A shows that certain morphological features (e.g., QRS amplitude, PR interval) exhibit deviations after decryption, highlighting a limitation that will be addressed in future refinements. Future work will explore tailoring chaotic systems to different signal types, integrating hardware accelerators (e.g., CNN or FPGA) for faster encryption, and developing secure cloud architectures for real-time encrypted storage, processing, and analysis.